\documentclass[twocolumn,prl,showpacs,preprintnumbers]{revtex4}

\usepackage{amsmath}
\usepackage{epsfig}
\usepackage{dcolumn}
\usepackage{bm}

\begin{document}

\title{Universal Fault-Tolerant Quantum Computation in the Presence of
Spontaneous Emission and Collective Dephasing}

\author{K. Khodjasteh L.}
\affiliation{Department of Physics, University of Toronto, 60
St.George St. Toronto, ON M5S 1A7, Canada.}
\author{D.A. Lidar}
\affiliation{Chemical Physics Theory Group, Chemistry Department,
University of Toronto, 80 St. George St., Toronto, ON M5S 3H6, Canada.}

\begin{abstract}
A universal and fault tolerant scheme for quantum computation is
proposed which utilizes a class of error correcting codes that is
based on the detection of spontaneous emission (of, e.g., photons,
phonons, and ripplons). The scheme is compatible with a number of
promising solid-state and quantum-optical proposals for quantum
computer implementations, such as quantum dots in cavities, electrons
on helium, and trapped ions.
\end{abstract}


\maketitle

The most severe obstacle in the path towards the dramatic speedup
offered by future quantum information processing (QIP) devices is
decoherence: the process whereby a quantum system becomes irreversibly
entangled with an uncontrollable environment (``bath''). This causes information loss and may degrade
the operation of a quantum computer to the point where it can be
efficiently simulated classically \cite{Aharonov:96a}. One can
formally model decoherence processes in QIP as being due to operators
$\{S_{i}\}$ acting on the system qubits $\{i\}$, that are coupled to
bath operators $B_{i}$ in a system-bath interaction Hamiltonian
$H_{SB}=\sum_{i}S_{i}\otimes B_{i}$. Two of the main proposals to
combat decoherence in QIP are \emph{quantum error correction codes}
(QECCs) and \emph{decoherence free subspaces} (DFSs). In QECCs multi-qubit states define quantum ``code words'', with the special property that they are
distinguishable (orthogonal) after the occurrence of errors, i.e.,
decoherence.  Appropriate non-destructive measurements yield an
``error syndrome'', which can be used
for recovery from the errors \cite{Knill:97,Gottesman:97a}. The DFS
approach similarly invokes code words, but it does not require active
measurement and recovery, since the encoded states are chosen so as to
be immune from decoherence:\ a state $|\psi _{n}\rangle $ is
decoherence-free if $ S_{i}|\psi _{n}\rangle =c_{i}|\psi _{n}\rangle$, where $c_{i}$ is a scalar that \emph{does not depend on} $|\psi
_{n}\rangle $ \cite{Zanardi:97a,Lidar:PRL99}. This condition, which we
refer to as \emph{the DFS condition} below, assumes that there is a
\emph{symmetry} in the system-bath interaction, such as
``collective decoherence'', wherein
$H_{SB}$ is qubit-permutation-invariant \cite{Zanardi:97c,Duan:98}. A
number of studies have pointed out the advantages of combining the
QECC and DFS approaches \cite{Lidar:PRL99,Lidar:00b,Alber:01}. Of
particular relevance is the recent work by Alber \textit{et al.} \cite{Alber:01}, who introduced a new class of hybrid DFS-QECC\ codes,
known as ``detected-jump correcting (DJC) quantum
codes''. These codes, which we review below, are
particularly useful in the case of spontaneous emission errors: $
S_{i}=\vert 0\rangle _{i}\langle 1|$, where $\vert 1\rangle _{i}$
($\vert 0\rangle _{i}$) is the excited (ground) state of, e.g., an
atom $i$. The DJC codes improve upon earlier work on QECC in the
presence of spontaneous emission \cite{Plenio:97a} in that they take
advantage of knowing where the emission event occurred (which
qubit). This assumes that the mean distance between qubits exceeds the
wavelength of the emission. The work by Alber \textit{et al.}
\cite{Alber:01} left open the question of computation with these codes
\cite{note}.

We show here how to perform universal, fault-tolerant quantum
computation (QC) on a class of the DJC hybrid codes, in the presence
of spontaneous emission and collective dephasing errors.  The latter
are errors that arise when the system-bath interaction can be written
as $H_{SB}=S_{z}\otimes B_{z}$, where $S_{z}=\sum_{i}\sigma _{i}^{z}$,
($\sigma _{i}^{z}$ is the Pauli $\sigma ^{z} $ matrix acting on the
$i^{\mathrm{th}}$ qubit), and have been extensively discussed before,
both theoretically
\cite{Duan:98,Lidar:PRL99,Bacon:99aKempe:00,WuLidar:01b} and
experimentally \cite{Kwiat:00,Kielpinski:01}. We show below that in
order to accomplish this \emph{we need only control the coupling constants
$J_{ij}^{z}$ and/or $J_{ij}$ appearing in an anisotropic,
exchange-type system Hamiltonian}: $H_{S}=\sum J_{ij}(\sigma
_{i}^{x}\sigma _{j}^{x}+\sigma _{i}^{y}\sigma
_{j}^{y})+J_{ij}^{z}\sigma _{i}^{z}\sigma _{j}^{z}$. The case
$J_{ij}^{z}\neq 0$ ($J_{ij}^{z}=0$) is known as the XXZ (XY)
model. These types of Hamiltonians naturally appear in a number of
promising proposals for implementing quantum computers, in which
spontaneous emission, as well as collective dephasing errors, are
important sources of decoherence. E.g., the quantum Hall
\cite{Mozyrsky:01}, quantum dots \cite{Imamoglu:99}, dimer
atoms in a solid host \cite{Petrosyan:02}, and atoms in cavities
\cite{Zheng:00} proposals are all of the XY type and suffer from
photon and phonon emission, while the electrons on helium proposal
\cite{Platzman:99} is of the XXZ type and suffers in addition from ripplon emission. The phonon-mediated ion-ion interaction in the
S$\o$rensen-M$\o$lmer (SM) scheme for trapped-ion QC \cite{Sorensen:99} is equivalent to an XY model, and this proposal too suffers from spontaneous emission
of photons and phonons, as well as from collective dephasing
\cite{Kielpinski:01}. Other sources of decoherence can also appear in
all proposals, but as shown in \cite{WuLidar:01b}, using appropriate
pulse sequences generated by the XY Hamiltonian, they can be reduced
to the collective dephasing type. The idea of universal QC using the
XY or XXZ interaction has been considered before, starting with
\cite{Imamoglu:99}, where the XY interaction had to be supplemented
with arbitrary single-qubit operations. In \cite{LidarWu:01} it was
shown how to perform universal QC using the XY interaction
supplemented with static single-qubit energy terms (e.g., a Zeeman
splitting) and an encoding into 2 qubits; in \cite{Kempe:01} universal
gate sequences were given for the XY interaction alone, using an
encoding into 3 qubits; and in \cite{LidarWu:01,WuLidar:01a} universal
gate sequences using the XXZ interaction were found for
encodings into 2 or more qubits. Here we use an encoding into 4 or
more qubits, that has the additional, significant advantage of
offering protection (using a QECC) against spontaneous emission
errors.

\textit{Detected Jump-Corrected Codes}.--- In the DJC codes method,
the Markovian quantum trajectories approach \cite{Plenio:98} is used
to describe decoherence. This approach is equivalent to the Lindblad
semigroup master equation \cite{Lindblad:76}. The evolution is
decomposed into two parts: a conditional \emph{non-Hermitian}
Hamiltonian $H_{C}$, interrupted at random times by application of
random errors. For errors such as spontaneous emission, where the jump
can be detected by observation of the emission, the quantum
trajectories approach also provides a way to combine QECCs and DFSs
\cite{Alber:01}. The DFS takes care of the conditional evolution,
whereas the QECC deals with the random jumps that couple DFS states
with states outside of the DFS. Formally, the conditional Hamiltonian
is given by \cite{Plenio:98}: $H_{C}=H_{S}-\frac{ i}{2}\sum_{i}\kappa
_{i}S_{i}^{\dag }S_{i}$, where $\kappa _{i}$ are (in our case) the
spontaneous emission rates. The DFS in the quantum jump approach is
given by the eigenspace of the \emph{collective} operator $C\equiv
\sum_{i}\kappa _{i}S_{i}^{\dag }S_{i}$. The symmetry that leads to the
DFS condition being satisfied is $\kappa _{i}\equiv \kappa $. For $n$
qubits and spontaneous emission errors we then have $C=\kappa
\sum_{i=1}^{n}\vert 1 \rangle_{i} \langle 1\vert $, and the DFS with
maximal dimension is comprised of (``computational'') basis states with $\frac{n}{2}$ $1$'s
and $\frac{n}{2}$ $0$'s. It has dimension $\binom{n}{n/2}$ and
eigenvalue $n/2$ under $C$. From here on we work exclusively with this
DFS.

Consider such a DFS encoding into $n=4$ qubits ($n=2$ qubits already
yields a logical qubit, but $n=4$ is the smallest such
\emph{generalizable} example, in the sense of the multi-encoded-qubit
scheme discussed below). It protects against
the conditional evolution, so what remains is to protect against the
jumps. As shown in \cite{Alber:01}, if we assume knowledge of the
position of errors by observing the emission, then one can use states
in this DFS in order to construct a QECC that encodes one logical
qubit:
\begin{eqnarray}
\vert 0\rangle _{L} =\frac{\vert 1010\rangle +\vert
0101\rangle}{\sqrt{2}} \quad \vert 1\rangle _{L} =\frac{\pm (\vert
0110\rangle +\vert 1001\rangle )}{\sqrt{2}}, \label{eq:4code}
\end{eqnarray}
where the choice of sign is $+$ ($-$) if $J_{12}<0$ (\thinspace $>0$),
as will be clarified below, and for simplicity we assume from here on
that $ J_{ij}^{z}\geq 0$. The general QECC condition \cite{Knill:97}
that keeps the errors from scrambling the code words $\vert \psi
_{n}\rangle $ takes the following form, provided we know which of the
errors indexed by $i$ has occurred:
\begin{equation}
\langle \psi _{m}\vert S_{i}^{\dag }S_{i}\vert \psi _{n}\rangle
=\Lambda _{i}\delta _{mn}, \label{eq:QECCcond}
\end{equation}
where $\Lambda _{i}$ is a number independent of the code words
\cite{Alber:01}. This is easily verified for the code in
Eq.~(\ref{eq:4code}).  Therefore this code offers complete protection
against the detected-jump spontaneous emission process. Note that in
addition to the states in Eq.~(\ref{eq:4code}) the state $\vert
2_{L}\rangle =(\vert 0011\rangle +\vert 1100\rangle )/\sqrt{2}$ also
satisfies the QECC condition (\ref{eq:QECCcond}), and is inside the
DFS.

Alber \textit{et al.} \cite{Alber:01} gave a combinatorial
design-theory method for generalizing the code of
Eq.~(\ref{eq:4code}). We now describe a class of these codes that come
with natural encoded qubit operations, that allow for \emph{universal,
scalable, and fault tolerant QC}. Our \emph{protocol} is as follows:
Computation is performed during the conditional evolution periods,
while the system is in a DFS. If a jump is detected, it must first be
corrected (as in QECC), before computation can resume. We note that the performance
of DJC codes in the presence of imperfections such as detection
inefficiencies, unequal decay rates $\kappa_i$, and time delay between
error detection and recovery operations, has been analyzed in
\cite{Alber:01}(c), with favorable conclusions.

\textit{Example: Universal Encoded Logic Operations for the 4-Qubit
DJC Code.--- } In order to perform universal QC we first identify a
set of generators of all encoded single qubit transformations. As is
well known, arbitrary single qubit transformations can be generated
from Hamiltonians via time evolution, using a standard Euler angle
construction: $e^{-i\omega \mathbf{n}\cdot
\boldsymbol{\sigma}}=e^{-i\beta \sigma ^{z}}e^{-i\theta \sigma
^{x}}e^{-i\alpha \sigma ^{z}}$. This is a rotation by angle $\omega $
about the axis $\mathbf{n}$, given in terms of three successive
rotations about the $z$ and $x$ axes. Let us now suppose that we have
at our disposal a controllable XXZ Hamiltonian, as defined above.
This gives us the ability to switch on/off, separately, the
Hamiltonian terms $T_{ij}\equiv \frac{1}{2} (X_{i}X_{j}+Y_{i}Y_{j})$
and $Z_{i}Z_{j}$, where $X_i \equiv \sigma^x_i$, etc. These operators
preserve the number of $0$'s and $1$'s
\cite{WuLidar:01,WuLidar:01a}. Since this implies that they cannot
take states outside of the DFS, it follows that they are
\emph{naturally fault tolerant} \cite{Bacon:99aKempe:00}.  Now suppose
that we turn $|J_{12}|$ ($J_{13}^{z}$) on for a time $t$ such that $
|J_{12}|t/\hbar =\theta $ ($J_{13}^{z}t/\hbar =\theta $). Then:
\begin{eqnarray}
e^{-i\theta T_{12}}\vert \epsilon \rangle _{L} &=&\cos \theta \vert
\epsilon \rangle _{L} -i\sin \theta \vert \bar{\epsilon} \rangle _{L},
\notag \\ e^{-i\theta Z_{1}Z_{3}}\vert 0\rangle _{L} &=&e^{-i\theta
}\vert 0\rangle _{L},\quad e^{i\theta Z_{1}Z_{3}}\vert 1\rangle
_{L}=e^{i\theta }\vert 1\rangle _{L}
\label{eq:XZn=4}
\end{eqnarray}
where $\epsilon =0$ or $1$, and $\bar{\epsilon}=(\epsilon+1)\mod
2=$\textsc{NOT}$(\epsilon )$. These equations show that $T_{12}$ and
$Z_{1}Z_{3}$ have precisely the action of single qubit $\sigma ^{x}$
and $\sigma ^{z}$ transformations, on the code states in
Eq.~(\ref{eq:4code}), and that this code space is perfectly preserved
under $T_{12}$ and $Z_{1}Z_{3}$. We denote logical $X$ ($ Z$)\
operations on the $i^{\mathrm{th}}$ encoded qubit by $\bar{X}_{i}$ ($
\bar{Z}_{i}$). Thus $\bar{X}_{1}=T_{12}$ and $\bar{Z}_{1}=Z_{1}Z_{3}$
and we have the ability to generate arbitrary encoded single qubit
transformations in the XXZ model. This is particularly relevant for
the electrons on helium proposal \cite{Platzman:99}.

However, in many QC proposals of interest it is either inconvenient to
separately control $J_{ij}^{z}$, or such exchange interactions vanish
\cite{Mozyrsky:01,Imamoglu:99,Petrosyan:02,Zheng:00,Sorensen:99}. We
must then resort to controlling only the XY term. Now, as shown in
\cite{LidarWu:01}, using the ``encoded
recoupling'' method, it is possible to generate
$Z_{2i-1}Z_{2j-1}$ operations with arbitrary $i,j$ as long as one can
control an XY Hamiltonian. Define $C_{A}^\phi \circ B\equiv \exp
(-i\phi A)B\exp (i\phi A)$, then \cite{WuLidar:01a,LidarWu:01,note2}:
\begin{equation}
2C_{\frac{1}{2}T_{2i,2j-1}}^{\pi/2}\circ (C_{T_{2i-1,2i}}^{\pi/2}\circ
T_{2i-1,2j-1}) = Z_{2i-1}Z_{2j-1}.  \label{eq:C}
\end{equation}
The procedure given in Eq.~(\ref{eq:C}) is a 5-step implementation of
the Ising interaction $Z_{2i-1}Z_{2j-1}$. For $i=1,j=2$ this yields $
\bar{Z}_{1}$, and we have all we need for encoded single qubit
transformations in the XY model.

The one apparent disadvantage of the procedure in Eq.~(\ref{eq:C}) is
that in 1D it requires next-nearest neighbor interactions (this is
inevitable with an XY interaction in 1D \cite{WuLidar:01a}), but note
that these interactions are still nearest neighbor on a 2D triangular
qubit lattice. Let us also note that application of $T_{2i-1,2j-1}$
[as arises in Eq.~(\ref{eq:C}); e.g., $T_{13}$ is needed for the
implementation of $\bar{Z}_{1}$], maps the code state $\vert 1\rangle
_{L}$ to a superposition of $ \vert 1\rangle _{L}$ and $\vert 2\rangle
_{L}$. While $ \vert 2\rangle _{L}$ is not part of our encoded qubit
it is part of the DJC code [it is in the DFS and satisfies the QECC
condition (\ref{eq:QECCcond})], so that the fault tolerance of our
procedure is not violated.

\textit{Generalization: DJC Code Encoding Several Qubits}.--- We now introduce an encoding that generalizes the code in
Eq.~(\ref{eq:4code}) to arbitrary numbers of encoded qubits. Let
\begin{equation*}
|\tilde{0}\rangle _{i}\equiv |0_{2i-1}1_{2i}\rangle ,\quad
|\tilde{1}\rangle _{i}\equiv
-\mathrm{sign}(J_{2i-1,2i})|1_{2i-1}0_{2i}\rangle .
\end{equation*}
We then define a code as follows:
\begin{equation}
|\epsilon _{L}\rangle _{1}\otimes \cdots \otimes |\epsilon _{L}\rangle
_{n-1}= \frac{|\tilde{\epsilon}\rangle _{1}\cdots
|\tilde{\epsilon}\rangle _{n-1}|\tilde{0}\rangle
_{n}+\mathrm{conj.}}{\sqrt{2}},
\label{eq:defcode}
\end{equation}
where $\epsilon =0$ or $1$, and ``$\mathrm{conj.}$'' denotes the bitwise \textsc{NOT} of the first
ket. The rate (number of encoded per physical qubits) of this class of
codes is $r= \frac{n-1}{2n}$. As in the case of a single encoded
qubit, Eq.~(\ref{eq:XZn=4}), the generators of encoded $\sigma ^{x}$
and $\sigma ^{z}$ transformations are
\begin{eqnarray}
\bar{X}_{i}=\frac{1}{2}(X_{2i-1}X_{2i}+Y_{2i-1}Y_{2i}),\quad
\bar{Z}_{i}= Z_{2i-1}Z_{2n-1},
\label{eq:XiZi}
\end{eqnarray}
as is easily verified by checking their action on $|\tilde{0}\rangle
_{i},| \tilde{1}\rangle _{i}$. Using the Euler angle formula we may
construct arbitrary encoded single-qubit operations from $\bar{X}_{i}$
and $\bar{Z} _{i} $, using operations from within the XY or XXZ models
only. The fact that we can apply such single encoded qubit operations
on the code in Eq.~(\ref{eq:defcode}) shows that \emph{this code is
equipped with a (formal) tensor product structure, and allows for
scalable QC}.

At this point we are ready to show how to implement a controlled-phase
(CP) gate, $\mathrm{CP}|x,y\rangle =\left( -1\right) ^{xy}|x,y\rangle
$ (where $ x,y$ are $0$ or $1$), which together with arbitrary
single-qubit operations is universal for QC \cite{Nielsen:book}.  As
is well known, the CP gate is generated by an Ising interaction
$Z\otimes Z$ \cite{Nielsen:book}. Thus to generate a CP gate between
\emph{encoded} qubits $i,j$ we must consider $
\bar{Z}_{i}\bar{Z}_{j}=\left( Z_{2i-1}Z_{2n-1}\right) \left(
Z_{2j-1}Z_{2n-1}\right) =Z_{2i-1}Z_{2j-1}$. In the XXZ model such a
two-body Ising interaction is directly controllable. In the XY model,
we can generate it using the 5-step procedure of
Eq.~(\ref{eq:C}). Furthermore, since a CP gate can be used to
construct a \textsc{SWAP} gate \cite{Nielsen:book}, we need only use
at most next nearest-neighbor interactions (in 1D; nearest neighbor in
2D) in order to couple arbitrary pairs of encoded qubits. Finally, we
stress that the combination of
Eqs.~(\ref{eq:XZn=4}),(\ref{eq:C}),(\ref{eq:XiZi}), and the
result above for $\bar{Z}_{i}\bar{Z}_{j}$, is an explicit prescription
for constructing arbitrary quantum circuits in terms of the XY and/or
XXZ interactions.

\textit{Fault Tolerant Measurement and Recovery}.--- An inherent
assumption in the DJC\ codes method is that it is possible to observe
which of the physical qubits underwent spontaneous emission
\cite{Alber:01}. This is a manifestly fault-tolerant measurement
\cite{Gottesman:97a}, in the sense that observing an error on a
specific qubit cannot cause errors to multiply. Now consider recovery
from spontaneous emission errors. If the error affects qubit $2i-1$
($2i$), the effect is $\vert \tilde{0}\rangle _{i}\mapsto \vert
\tilde{0}\rangle _{i}$, $\vert \tilde{1}\rangle _{i}=\vert
0_{2i-1}0_{2i}\rangle $ ($\vert \tilde{0}\rangle _{i}\mapsto \vert
0_{2i-1}0_{2i}\rangle $, $\vert \tilde{1}\rangle _{i}=\vert
\tilde{1}\rangle _{i}$). The recovery operation must therefore
correspondingly take $\vert 0_{2i-1}0_{2i}\rangle $ to $ \vert
\tilde{1}\rangle _{i}$ ($\vert \tilde{0}\rangle _{i}$), while not
affecting $\vert \tilde{0}\rangle _{i}$ ($\vert \tilde{1}\rangle_{i}$).  Corresponding unitary operations with the desired effect are
cousins of the standard controlled-NOT
\cite{Nielsen:book}, defined on the subspace of qubits
$2i-1,2i$:
\begin{equation*}
\text{\textsc{CX}}_{1}=\left(
\begin{array}{cccc}
& & 1 & \\ & 1 & & \\ 1 & & & \\ & & & 1
\end{array}
\right) ,\quad \text{\textsc{CX}}_{2}=\left(
\begin{array}{cccc}
& 1 & & \\ 1 & & & \\ & & 1 & \\ & & & 1
\end{array}
\right) .
\end{equation*}
Now, in order to perform these recovery operations we must assume that
in addition to an XY or XXZ Hamiltonian we have the ability to control
single-qubit energies (i.e., control terms of the form $\omega_i Z_i$)
and perform a Hadamard
[$W=\frac{1}{\sqrt{2}} \left(
\begin{array}{cc}
1 & 1 \\ 1 & -1
\end{array}
\right) $] gate, which is certainly reasonable in optics-based
QC proposals \cite{Imamoglu:99,Petrosyan:02,Zheng:00,Sorensen:99} (where
such single-qubit operations are executed through the application of
laser pulses). This requirement is harder to satisfy in solid-state QC
proposals that use gate voltages for single qubit operations
\cite{Mozyrsky:01,Platzman:99}, but is not
unreasonable. Note that the assumption that we can perform single
qubit operations is made only to enable recovery from spontaneous emission
errors. It is needed since the XY and XXZ Hamiltonians preserve the
number of $0$'s and $1$'s in each codeword, while spontaneous emission
lowers the number of $ 1$'s. Now, with the extra assumption we are
able to construct \textsc{CX}$ _{1}$ and \textsc{CX}$_{2}$ in 7 (3)
steps, assuming a controllable XY (XXZ) Hamiltonian. E.g.,
\textsc{CX}$_{1}=e^{-i\frac{\pi}{4}} (WP \otimes P^2)\exp (i\frac{\pi
}{4}Z_{1}Z_{2})(W\otimes P)$, where $P\equiv
e^{-i(3\pi)/4) Z} = e^{i(3\pi/4)}{\rm diag}(i,1)$, and we recall
that 5 steps are needed to generate $Z_{1}Z_{2}$ from the XY
Hamiltonian. To obtain \textsc{CX}$_{2}$ swap the order of the factors
around the $\otimes$ symbols. Since we apply \textsc{CX}$_{1}$ and \textsc{CX}$_{2}$ only
within a block encoding a single qubit the operations we perform can
only affect that encoded qubit.  Therefore if the operations
themselves are faulty the error cannot spread to other encoded
qubits. This means that our recovery operations are fault-tolerant
\cite{Gottesman:97a}.

\textit{State Preparation and Read-Out}.--- Finally, we must also show
that our encoded states can be reliably prepared and read out. A
general preparation technique is cooling to the ground state of a
Hamiltonian. For this procedure to work there should be an energy gap
$\Delta$ between the code subspace and other states. Diagonalization
of the XY Hamiltonian $
J_{ij}T_{ij}=\frac{J_{ij}}{2}(X_{i}X_{j}+Y_{i}Y_{j})$ in the subspace
of qubits $i,j$ yields, depending on whether $J_{ij}>0$ or $<0$,
either the singlet state $|s\rangle
_{ij}=\frac{1}{\sqrt{2}}(|0_{i}1_{j}\rangle -|1_{i}0_{j}\rangle )$ or
the triplet state $|t\rangle _{ij}=\frac{1}{\sqrt{2
}}(|0_{i}1_{j}\rangle +|1_{i}0_{j}\rangle )$, as the ground state,
with energy $-|J_{ij}|$. Consider the case of a single encoded qubit
and assume $ J_{ij}>0$:\ the ground state of the XY\ Hamiltonian $
J_{12}T_{12}+J_{34}T_{34}$ is $|s_{12}\rangle \otimes |s_{34}\rangle
$, which is exactly $\frac{1}{\sqrt{2}}\left( \vert 0\rangle
_{L}+\vert 1\rangle _{L}\right) $, in terms of the code states of
Eq.~(\ref{eq:4code}) with the choice of ``$-$'' for $\vert 1\rangle
_{L}$. I.e., cooling prepares a state that is in the code subspace,
and application of the encoded logical operations derived above can
rotate this initial state to any other desired encoded state. To
prepare a state in the code subspace of $2n$ physical qubits we turn
on the pairwise XY Hamiltonian $\sum_{i=1}^{n}J_{2i-1,2i}T_{2i-1,2i}$,
keep the temperature below $\Delta$, and wait. The resulting ground
state is $\otimes _{i=1}^{n}|s\rangle _{2i-1,2i}$, and a simple
calculation shows that this state is in the code space:
\begin{eqnarray*}
\otimes _{i=1}^{n}|s\rangle _{2i-1,2i} = \otimes _{j=1}^{n-1}\left(
|0_{L}\rangle _{j}+|1_{L}\rangle _{j}\right) / \sqrt{2}.
\end{eqnarray*}
Identical conclusions hold when assuming $J_{ij}<0$, with $|t\rangle
_{2i-1,2i}$ replacing $|s\rangle _{2i-1,2i}$. Thus cooling always
prepares a state in the code subspace and can serve as an
initialization procedure for our protocol. Measurement can be be done
analogously, i.e., by using the energy difference to distinguish a
singlet from a triplet state on pairs of qubits encoding a logical
qubit \cite{WuLidar:01}. Thus, to distinguish $ |0_{L}\rangle _{j}$ from
$|1_{L}\rangle _{j}$ we first apply an encoded Hadamard gate to
physical qubits $2j-1,2j$, mapping $|0_{L}\rangle _{j}\rightarrow
\left( |0_{L}\rangle _{j}+|1_{L}\rangle _{j}\right) /\sqrt{2} $ and
$|1_{L}\rangle _{j}\rightarrow \left( |0_{L}\rangle _{j}-|1_{L}\rangle
_{j}\right) /\sqrt{2}$, which by the preparation arguments above
correspond to singlet and triplet states, depending on
$\mathrm{sign}(J_{2j-1,2j})$.

\textit{Conclusions}.--- We have studied a class of ``detected-jump''
codes that is capable of avoiding collective dephasing errors and
correcting spontaneous emission errors on a single qubit. These codes
are a hybrid of decoherence-free subspaces and active quantum error
correction, and use $2n$ qubits to encode $n-1$. We have shown how to
quantum compute universally and fault tolerantly on this class of
codes, using Hamiltonians (XY- and XXZ-type) that are directly
relevant to a number of promising solid-state and quantum-optical
proposals for quantum computer implementations
\cite{Mozyrsky:01,Imamoglu:99,Petrosyan:02,Zheng:00,Platzman:99,Sorensen:99}.

\begin{acknowledgments}
We thank Dr. L.-A. Wu for very helpful discussions and Photonics
Research Ontario for financial support (to D.A.L.).
\end{acknowledgments}


\begin{thebibliography}
\bibitem{} \expandafter\ifx\csname natexlab\endcsname\relax

\fi \expandafter\ifx\csname bibnamefont\endcsname\relax

\fi \expandafter\ifx\csname bibfnamefont\endcsname\relax

\fi \expandafter\ifx\csname citenamefont\endcsname\relax

\fi \expandafter\ifx\csname url\endcsname\relax

\fi \expandafter\ifx\csname urlprefix\endcsname\relax

\fi \providecommand{\bibinfo}[2]{#2} \providecommand{\eprint}[2][]{\url{#2}}

\bibitem[{D. Aharonov and M. Ben-Or}(1996)]{Aharonov:96a}
\bibinfo{author}{\bibnamefont{{D. Aharonov and M. Ben-Or}}}, in \emph{
\bibinfo{booktitle}{{Proceedings of 37th Conference on Foundations of
  Computer Science (FOCS)}}} (\bibinfo{publisher}{{IEEE
Comput. Soc. Press}}, \bibinfo{address}{{Los Alamitos, CA}},
\bibinfo{year}{1996}), p.~\bibinfo{pages}{46}.

\bibitem[{E. Knill and R. Laflamme}(1997)]{Knill:97}
\bibinfo{author}{\bibnamefont{{E. Knill and R. Laflamme}}},
\bibinfo{journal}{Phys. Rev. A} \textbf{\bibinfo{volume}{55}},
\bibinfo{pages}{900} (\bibinfo{year}{1997}).

\bibitem[Gottesman(1997)]{Gottesman:97a} \bibinfo{author}{\bibfnamefont{D.}~
\bibnamefont{Gottesman}}, \bibinfo{journal}{Phys. Rev. A} \textbf{
\bibinfo{volume}{57}}, \bibinfo{pages}{127} (\bibinfo{year}{1997}).

\bibitem[{P. Zanardi, M. Rasetti}(1997{a})]{Zanardi:97a}
\bibinfo{author}{\bibnamefont{{P. Zanardi, M. Rasetti}}},
\bibinfo{journal}{Mod. Phys. Lett. B} \textbf{\bibinfo{volume}{11}},
\bibinfo{pages}{1085} (\bibinfo{year}{1997}{\natexlab{a}}).

\bibitem[{D.A. Lidar, D. Bacon and K.B. Whaley}(1999)]{Lidar:PRL99}
\bibinfo{author}{\bibnamefont{{D.A. Lidar {\it et al.}}}},
\bibinfo{journal}{Phys. Rev. Lett.} \textbf{\bibinfo{volume}{82}},
\bibinfo{pages}{4556} (\bibinfo{year}{1999}).

\bibitem[{P. Zanardi, M. Rasetti}(1997{b})]{Zanardi:97c}
\bibinfo{author}{\bibnamefont{{P. Zanardi, M. Rasetti}}},
\bibinfo{journal}{Phys. Rev. Lett.} \textbf{\bibinfo{volume}{79}},
\bibinfo{pages}{3306} (\bibinfo{year}{1997}{\natexlab{b}});
\bibinfo{author}{\bibnamefont{{D.A. Lidar {\it et al.}}}},
\bibinfo{journal}{Phys. Rev. Lett.} \textbf{\bibinfo{volume}{81}},
\bibinfo{pages}{2594} (\bibinfo{year}{1998}).

\bibitem{Duan:98} \bibinfo{author}{\bibnamefont{{L.-M Duan and G.-C. Guo}}},
\bibinfo{journal}{Phys. Rev. A} \textbf{\bibinfo{volume}{57}},
\bibinfo{pages}{737} (\bibinfo{year}{1998}).

\bibitem[{D.A. Lidar, D. Bacon, J. Kempe, and K.B. Whaley}(2001)]{Lidar:00b}
\bibinfo{author}{\bibnamefont{{D.A. Lidar {\it et al.}}}},
\bibinfo{journal}{Phys. Rev. A} \textbf{\bibinfo{volume}{63}},
\bibinfo{pages}{022307} (\bibinfo{year}{2001}).

\bibitem[{G. Alber, Th. Beth, Ch. Charnes, A. Delgado, M. Grassl, and M.
Mussinger}(2001)]{Alber:01} (a) \bibinfo{author}{\bibnamefont{{G. Alber {\it et al.}}}},
\bibinfo{journal}{Phys. Rev. Lett.} \textbf{\bibinfo{volume}{86}},
\bibinfo{pages}{4402} (\bibinfo{year}{2001}); (b)
\bibinfo{journal}{Fortschr. Phys.} \textbf{\bibinfo{volume}{49}},
\bibinfo{pages}{901} (\bibinfo{year}{2001}); (c) \eprint{eprint quant-ph/0208140}.

\bibitem[{M.B. Plenio, V. Vedral, and P.L. Knight}(1997)]{Plenio:97a}
\bibinfo{author}{\bibnamefont{{M.B. Plenio {\it et al.}}}},
\bibinfo{journal}{Phys. Rev. A} \textbf{\bibinfo{volume}{55}},
\bibinfo{pages}{67} (\bibinfo{year}{1997}).

\bibitem{note}
{Except a brief discussion in \cite{Alber:01}(b), where
it was shown that the \emph{Heisenberg} exchange interaction can be
used to perform single qutrit operations for one instance of the DJC
code. See also A. Beige {\it et al.}, Phys. Rev. Lett. \textbf{85},
1762 (2000).}

\bibitem[{D. Bacon, J. Kempe, D.A. Lidar and K.B. Whaley}(2000)]{Bacon:99aKempe:00}
\bibinfo{author}{\bibnamefont{{D. Bacon {\it et al.}}}},
\bibinfo{journal}{Phys. Rev. Lett.} \textbf{\bibinfo{volume}{85}},
\bibinfo{pages}{1758} (\bibinfo{year}{2000});
\bibinfo{author}{\bibnamefont{{J. Kempe {\it et al.}}}},
\bibinfo{journal}{Phys. Rev. A} \textbf{\bibinfo{volume}{63}},
\bibinfo{pages}{042307} (\bibinfo{year}{2001}).

\bibitem[{L.-A. Wu and D.A. Lidar}(2002{a})]{WuLidar:01b}
\bibinfo{author}{\bibnamefont{{L.-A. Wu and D.A. Lidar}}},
\bibinfo{journal}{Phys. Rev. Lett.} \textbf{\bibinfo{volume}{88}},
\bibinfo{pages}{207902} (\bibinfo{year}{2002});
\bibinfo{author}{\bibnamefont{L. Viola}},
\bibinfo{journal}{Phys. Rev. A} \textbf{\bibinfo{volume}{66}},
\bibinfo{pages}{012307} (\bibinfo{year}{2002}).


\bibitem[{P.G. Kwiat, A.J. Berglund, J.B. Altepeter, and A.G. White}(2000)]
{Kwiat:00} \bibinfo{author}{\bibnamefont{{P.G. Kwiat {\it et al.}}}},
\bibinfo{journal}{Science} \textbf{\bibinfo{volume}{290}},
\bibinfo{pages}{498} (\bibinfo{year}{2000}).

\bibitem[{D. Kielpinski, V. Meyer, M. A. Rowe, C. A. Sackett,
W. M. Itano, C. Monroe, and D. J. Wineland}(2001)]{Kielpinski:01}
\bibinfo{author}{\bibnamefont{{D. Kielpinski {\it et al.}}}},
\bibinfo{journal}{Science}
\textbf{\bibinfo{volume}{291}}, \bibinfo{pages}{1013} (\bibinfo{year}{2001}).

\bibitem[{D. Mozyrsky, V. Privman, and M.L. Glasser}(2001)]{Mozyrsky:01}
\bibinfo{author}{\bibnamefont{{D. Mozyrsky {\it et al.}}}},
\bibinfo{journal}{Phys. Rev. Lett.} \textbf{\bibinfo{volume}{86}},
\bibinfo{pages}{5112} (\bibinfo{year}{2001}).

\bibitem[{A. Imamo$\bar{\mathrm{g}}$lu, D.D. Awschalom, G. Burkard, D.P.
DiVincenzo, D. Loss, M. Sherwin and A. Small}(1999)]{Imamoglu:99}
\bibinfo{author}{\bibnamefont{{A. Imamo$\bar{\rm g}$lu {\it et al.}}}},
\bibinfo{journal}{Phys. Rev. Lett.} \textbf{\bibinfo{volume}{83}},
\bibinfo{pages}{4204} (\bibinfo{year}{1999}).

\bibitem[{D. Petrosyan and G. Kurizki}(2002)]{Petrosyan:02}
\bibinfo{author}{\bibnamefont{{D. Petrosyan and G. Kurizki}}},
\eprint{eprint quant-ph/0205174}.

\bibitem[{S.-B. Zheng and G.-C Guo}(2000)]{Zheng:00} \bibinfo{author}{
\bibnamefont{{S.-B. Zheng and G.-C Guo}}},
\bibinfo{journal}{Phys. Rev. Lett.} \textbf{\bibinfo{volume}{85}}, \bibinfo{pages}{2392} (\bibinfo{year}{2000}).

\bibitem[{P.M. Platzman and M.I. Dykman}(1999)]{Platzman:99}
\bibinfo{author}{\bibnamefont{{P.M. Platzman and M.I. Dykman}}},
\bibinfo{journal}{Science} \textbf{\bibinfo{volume}{284}},
\bibinfo{pages}{1967} (\bibinfo{year}{1999}).

\bibitem[{A. Sorensen and K. Molmer}(1999)]{Sorensen:99}
\bibinfo{author}{\bibnamefont{{A. S$\o$rensen and K. M$\o$lmer}}},
\bibinfo{journal}{Phys. Rev. Lett.} \textbf{\bibinfo{volume}{82}},
\bibinfo{pages}{1971} (\bibinfo{year}{1999}).

\bibitem[{D.A. Lidar and L.-A. Wu}(2002)]{LidarWu:01} \bibinfo{author}{
\bibnamefont{{D.A. Lidar and L.-A. Wu}}}, \bibinfo{journal}{Phys. Rev. Lett.}
\textbf{\bibinfo{volume}{88}}, \bibinfo{pages}{017905} (\bibinfo{year}{2002}).

\bibitem[{J. Kempe, D. Bacon, D.P. DiVincenzo and K.B. Whaley}(2001)]
{Kempe:01} \bibinfo{author}{\bibnamefont{{J. Kempe {\it et al.}}}},
\bibinfo{journal}{Quant. Inf. Comp.} \textbf{\bibinfo{volume}{1}},
\bibinfo{pages}{33} (\bibinfo{year}{2001}).

\bibitem[{L.-A. Wu and D.A. Lidar}(2002{b})]{WuLidar:01a}
\bibinfo{author}{\bibnamefont{{L.-A. Wu and D.A. Lidar}}},
\bibinfo{journal}{J. Math. Phys.} \textbf{\bibinfo{volume}{43}},
\bibinfo{pages}{4506} (\bibinfo{year}{2002}).

\bibitem[{M. Plenio and P. Knight}(1998)]{Plenio:98}
\bibinfo{author}{\bibnamefont{{M. Plenio and P.
Knight}}}, \bibinfo{journal}{Rev. Mod. Phys.} \textbf{\bibinfo{volume}{70}},
\bibinfo{pages}{101} (\bibinfo{year}{1998}).

\bibitem[{G. Lindblad}(1976)]{Lindblad:76}
\bibinfo{author}{\bibnamefont{{G. Lindblad}}}, \bibinfo{journal}{Commun. Math.
  Phys.} \textbf{\bibinfo{volume}{48}}, \bibinfo{pages}{119}
(\bibinfo{year}{1976}).

\bibitem[{L.-A. Wu and D.A. Lidar}(2002{c})]{WuLidar:01} \bibinfo{author}{
\bibnamefont{{L.-A. Wu and D.A. Lidar}}}, \bibinfo{journal}{Phys. Rev. A}
\textbf{\bibinfo{volume}{65}}, \bibinfo{pages}{042318}
(\bibinfo{year}{2002}{\natexlab{c}}).

\bibitem{note2}
{Here we have neglected a term $-Z_{2i-1}Z_{2i}/2$ since it is constant on the
code space, i.e., it has equal action on $|0_{2i-1}1_{2i}\rangle
,|1_{2i-1}0_{2i}\rangle $.}

\bibitem[{M.A. Nielsen and I.L. Chuang}(2000)]{Nielsen:book}
\bibinfo{author}{\bibnamefont{{M.A. Nielsen and I.L. Chuang}}}, \emph{
\bibinfo{title}{{Quantum Computation and Quantum Information}}} (\bibinfo{publisher}{{Cambridge University Press}},
\bibinfo{address}{Cambridge, UK}, \bibinfo{year}{2000}).

\end{thebibliography}
\end{document}